\documentclass[aps,prd,twocolumn,floatfix,showpacs,nofootinbib]{revtex4} 

\usepackage{graphics}
\usepackage{graphicx}

\usepackage{latexsym}
\usepackage[cp866]{inputenc}
\usepackage[english]{babel}  

\input epsf

\begin{document}

\title{\boldmath $f_0(980)$ and $a_0(980)$ resonances near
$\gamma\gamma\to K^+K^-$ and $\gamma\gamma\to K^0\bar K^0$\\
reaction thresholds}
\author{N.N. Achasov\footnote{E-mail: achasov@math.nsc.ru}
and G.N. Shestakov\footnote{E-mail: shestako@math.nsc.ru}
}\affiliation{Laboratory of  Theoretical Physics, S.L. Sobolev
Institute for Mathematics, 630090, Novosibirsk, Russia}

\begin{abstract}
High-statistics data on the reactions $\gamma\gamma\to K^+K^-$ and
$\gamma\gamma\to K^0\bar K^0$ are the last missing link in
investigations of the light scalar mesons $f_0(980)$ and $a_0(980)$
in photon-photon collisions. It is believed that $f_0(980)$ and
$a_0(980)$ resonances exhibit their four-quark structure in these
reactions in a vary peculiar way. The work estimates the feasibility
of measurements of scalar contributions near $\gamma\gamma\to
K^+K^-$ and $\gamma\gamma\to K^0\bar K^0$ reaction thresholds at
modern colliders.\end{abstract}


\maketitle

A major contribution to understanding the nature of light scalar
mesons $\sigma(600)$, $f_0(980)$, and $a_0(980)$ (which are
candidates for four-quark states) has come from the physics of
photon-photon collisions, a field which has recently entered the era
of high-precision statistics (see, for example, a recent review
\cite{AS11}). It has been opened by the unprecedented series of
measurements of the $\gamma\gamma$\,$\to$\,$\pi^+\pi^-$
\cite{Mo07a,Mo07b}, $\gamma \gamma$\,$\to$\,$\pi^0\pi^0$
\cite{Ue08}, $\gamma\gamma$\,$\to $\,$\pi^0\eta$ \cite{Ue09}, and
$\gamma\gamma$\,$\to$\,$\eta\eta$ \cite{Ueh10} reaction cross
sections, performed by the Belle Collaboration at KEKB. The
statistics collected in these experiments is two to three order of
magnitude higher than in pre-$B$-factory measurements. Recently, the
two-photon production of the $\pi^0\eta$ system has been also
investigated by the BABAR Collaboration at PEP-II \cite{Sa11}.

Extensive programs on the two-photon physics, aimed, in particular,
at continuing precision measurements of the $\gamma\gamma$\,$\to
$\,$\pi^+ \pi^-$, $\gamma\gamma$\,$\to$\,$ \pi^0\pi^0$, and
$\gamma\gamma$\,$\to $\,$\pi^0\eta$ processes in the $\sigma(600)$,
$f_0(980)$, and $a_0(980)$ resonance region, are preparing for
realization at the upgraded collider BEPC\,II (with a luminosity of
$10^{33}$\,cm$^{-2}$sec$^{-1}$) with the use of the BES-III detector
\cite{As09} and  at the upgraded $\phi$-factory DA$\Phi$NE (with a
luminosity of $(1-5)\times10^{32}$\,cm$^{-2}$sec$ ^{-1}$) with the
use of the KLOE-2 detector \cite{Ve10,A-C10,Ba10, Cz10,Ar11}.

Similar two-photon experiments are also possible at the VEPP-4M
accelerator with the KEDR detector \cite{An10} and at the VEPP-2000
accelerator (with a luminosity of $10^{32}$\,cm$^{-2}$sec$^{-1}$)
with the detectors CMD-3 \cite{So11} and SND \cite{As11}.

High-statistic information is still lacking for the $\gamma\gamma
$\,$\to$\,$ K^+K^-$ and $\gamma\gamma$\,$\to$\,$ K^0\bar K^0$
processes in the energy range around 1 GeV. It is believed that
$f_0(980)$ and $a_0(980)$ resonances exhibit their four-quark
structure in these processes in a vary peculiar way
\cite{AS91,AS92,AS94b}.

Experiments show that the cross sections of  $\gamma\gamma
$\,$\to$\,$K^+K^-$ \cite{Al83,Jo86,Ai86,Alb90,FH91,Ab04} and
$\gamma\gamma $\,$\to$\,$K^0_SK^0_S$
\cite{Al83,FH91,Alt85,Ber88,Beh89,Bra00,Acc01,Hu01} reactions in an
energy range of $1.2<\sqrt{s}<1.7$ GeV ($\sqrt{s}$ is the invariant
mass of the $\gamma\gamma$ system) are actually saturated with the
contributions from classical tensor $f_2(1270)$, $a_2(1320)$, and
$f'_2(1525)$ resonances produced in helicity states with
$\lambda$\,=\,$\pm2$. Constructive and destructive interference
between $f_2(1270)$- and $a_2(1320)$-resonance contributions are
observed in $\gamma\gamma$\,$\to$\,$K^+K^-$ and
$\gamma\gamma$\,$\to$\,$K^0\bar K^0$ reactions, respectively, in
agreement with the $q\bar q$ model prediction \cite{FLR75}. The
energy region near the $K\bar K$ thresholds, $2m_K<\sqrt{s}<1.1$
GeV, sensitive to the $S$-wave contributions, remains virtually
unexplored. In the ARGUS experiment \cite{Alb90}, the efficacy of
recording $K^+K^-$ events for $2m_{K^+}<\sqrt{s}<1.1$ GeV was
negligible, while the statistics in the L3 experiment \cite{Acc01}
on the $\gamma\gamma $\,$\to$\,$K^0_SK^0_S$ reaction for
$2m_{K^0}<\sqrt{s}<1.1$ GeV did not exceed 10 events. The available
data from other experiments relate to the region of $\sqrt{s}>1.2$
GeV. Note that the tensor resonance contributions are strongly
suppressed for $2m_K<\sqrt{s}<1.1$ GeV due to the $D$-wave threshold
factor $p^5_K(s) $\,=\,$(s/4-m^2_K)^{5/2}$. A simple estimate shows
that the $\gamma\gamma$\,$\to$\,$ K^+K^-$ cross section,
corresponding to all tensor contributions (including the Born
contribution), makes $\approx[p^5_{K^+}(s)/p^5_{K^+}
(1.21\,\mbox{GeV}^2)]\times2$\,nb for $2m_{K^+}<\sqrt{s}<1.1$ GeV
\cite{Alb90}. The $\gamma \gamma$\,$\to$\,$K^0 \bar K^0$ cross
section in this region, caused by the tails of tensor mesons, is at
least twenty times smaller. Figure 1 illustrates the scale of the
$K^+K^-$ production cross section observed in $\gamma\gamma$
collisions in the tensor meson region.

\begin{figure}\hspace{1.5mm}
\includegraphics[width=19pc]{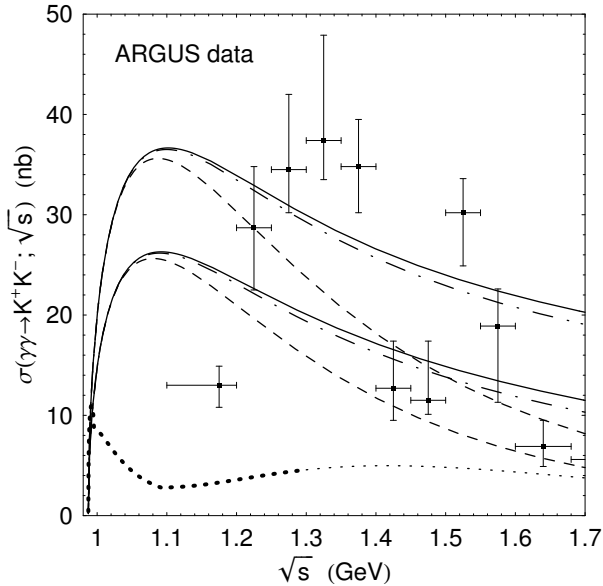}  
\caption{{\footnotesize Illustration of the scale of the $K^+K^-$
production cross section in $\gamma\gamma$ collisions. The data are
from ARGUS \cite{Alb90}. The upper dashed, dashed-dot, and solid
curves correspond to the Born $\gamma\gamma\to K^+K^-$ cross section
for the elementary (point-like) one-kaon exchange with $\lambda
J$\,=00, $|\lambda| J$\,=\,(00 and 22) and to the total Born cross
section, respectively (the Born contribution with $\lambda J$\,=02
is negligible); here $\lambda$ is the sum of helicities of initial
photons and $J$ is their total angular momentum. The lower dashed,
dashed-dot, and solid curves show the same cross section modified by
the form factor \cite{AS11,AS92,AS94b}. The dotted curve is our
estimate of the $S$-wave $\gamma\gamma\to K^+K^-$ cross section
\cite{AS11}.}}\label{gg-KK}\end{figure}

The absence of an appreciable nonresonant background in the $\gamma
\gamma$\,$ \to$\,$K^+K^-$ cross section seems at first sight rather
surprising, since the Born contribution mediated through the charged
one-kaon exchange mechanism and comparable with the tensor resonance
contributions must be present in this channel, see the solid curves
in Fig. 1. This figure also shows that the Born cross section [both
for the elementary (point-like) $K^\pm$ exchange and for the $K^\pm$
exchange with a form factor] is dominated by the $S$-wave
contribution for $2m_{K^+}<\sqrt{s}<1.5$ GeV. For this reason, a
large noncoherent background could be expected under tensor meson
peaks in the $K^+K^-$ channel. However, taking account of the
resonant interaction between $K^+$- and $K^-$-mesons in the final
state results in the compensation of a considerable part of this
background \cite{AS92,AS94b}. The compensation arises in the
following way. Due to the contribution from the $\gamma\gamma$\,$
\to$\,$K^+K^-$\,$ \to$\,$K^+K^-$ rescattering amplitude with real
kaons in the intermediate state, the Born $S$-wave $\gamma\gamma
$\,$\to$\,$K^+K^-$ amplitude acquires the factor $\xi(s)$\,=\,$[1
$\,+\,$i\rho_{K^+}(s)T_{K^+ K^-\to K^+K^-}(s)]$, where
$\rho_{K^+}(s)$\,=\,$2p_{K^+}(s)/\sqrt{s}$. Near the $K^+K^-$
threshold, the $S$-wave $T_{K^+K^-\to K^+K^-}(s)$ amplitude is
dominated by contributions from $f_0(980)$ and $a_0(980)$
resonances. Given their strong coupling to $K\bar K$-channels,
naturally realized in the four-quark scheme, the $T_{K^+K^-\to
K^+K^-}(s)$ amplitude possesses an appreciable imaginary part. As a
result, factor $|\xi(s)|^2$ just above the $K^+K^-$ threshold is
much smaller than unity and the seed $S$-wave Born contribution is
compensated for over a wide $\sqrt{s}$ range. The dotted curve in
Fig. 1 represents our estimate of the $S$-wave $\gamma\gamma\to
K^+K^-$ cross section \cite{AS11}, which fairly well agrees with the
ones obtained in earlier studies \cite{AS92,AS94b}. The validity of
these estimates can be expected at least for $\sqrt{s}\lesssim$\,1.3
GeV (see Fig. 1).

Thus, one can hope to detect scalar contributions at the level of
5--10 nb in the $\gamma\gamma$\,$\to$\,$K^+K^-$ cross section for
$2m_{K^+}<\sqrt{s}<1.1$ GeV. As regards the
$\gamma\gamma$\,$\to$\,$K^0\bar K^0$ reaction, its amplitude does
not contain the Born contribution, while the $a_0(980)$-resonance
contribution has the sign opposite to that in the $\gamma\gamma$\,$
\to$\,$K^+K^-$ channel. As a result, the contributions of $S$-wave
$\gamma\gamma$\,$\to$\,$K^+K^-$\,$ \to$\,$K^0\bar K^0$ rescattering
amplitudes with isospin $I$\,=\,0 and 1 in the $\gamma \gamma$\,$\to
$\,$K^0\bar K^0$ reaction practically cancel each other and the
corresponding cross section should be at the level of $\lesssim$\,1
nb. \footnote{Recall that the classical tensor $f_2(1270)$,
$a_2(1320)$, and $f'_2(1525)$ mesons couple to photons via direct
$q\bar q$\,$\to$\,$\gamma\gamma$ transitions, whereas the couplings
of the light scalar $\sigma(600)$, $f_0(980)$, and $a_0(980)$ mesons
to $\gamma\gamma$ are realized owing to the four-quark transitions
(rescattering mechanisms) of the type $\sigma
(600)$\,$\to$\,$\pi^+\pi^- $\,$\to$\,$\gamma\gamma$, $f_0(980)$\,$
\to$\,$K^+K^-$\,$\to $\,$\gamma\gamma$, $a_0(980)$\,$ \to$\,$(K^+
K^-,\pi^0\eta)$\,$\to$\,$\gamma\gamma$, and the direct
$\sigma(600)$\,$\to$\,$\gamma\gamma$, $f_0(980)$\,$\to$\,$
\gamma\gamma$, and $a_0(980)$\,$\to$\,$\gamma\gamma$ transitions are
small \cite{AS11}. Opportunity to explain the suppression of the
large $S$-wave Born contribution in $\gamma\gamma$\,$\to$\,$K^+K^-$
by that of the $f_0(980)$ and $a_0(980)$ resonances,
$\gamma\gamma$\,$\to$\,$ K^+K^-$\,$\to
$\,$[f_0(980)+a_0(980)]$\,$\to$\,$K^+K^-$, indicates in favor of
this picture and, consequently, in favor of the $q^2\bar q^2$ nature
of the $f_0(980)$ and $a_0(980)$ states.}

The number of two-photon events, $N_{eeX}$, produced in the
$e^+e^-$\,$\to$\,$e^+e^-X$ reaction, when $e^+$ and $e^-$ in the
final state are not registered, can be evaluated according to (see,
for example, Refs. \cite{As09,Ve10,A-C10})
\begin{equation}\label{NeeX}
N_{eeX}=L_{ee}\int\limits_{\Delta\sqrt{s}}
\frac{dF}{d\sqrt{s}}\sigma(\gamma\gamma\to X;\sqrt{s})d\sqrt{s}\,,
\end{equation} where $L_{ee}$ is the $e^+e^-$ integrated luminosity,
$\Delta\sqrt{s}$ is the interval of integration over the
$\gamma\gamma$ invariant mass, and $dF/d\sqrt{s}$ is the effective
$\gamma\gamma$ luminosity per unit energy,
\begin{equation}\label{Fgg}
\frac{dF}{d\sqrt{s}}=\frac{1}{\sqrt{s}}\left(\frac{2\alpha}
{\pi}\right)^2\left(\ln\frac{E_{cm}}{2m_e}\right)^2f(z)\,,
\end{equation} where $E_{cm}$ is the energy in the $e^+e^-$
center-of-mass system, $f(z)=-(z^2+2)^2\ln z-(1-z^2)(3+z^2)$, and
$z=\sqrt{s}/E_{cm}$.

Estimates of the number of events of two-photon production of
$K^+K^-$ pairs in the $S$-wave, $N_{eeK^+K^-}$, are presented in
Table I for working values of $E_{cm}$ and probable values of
$L_{ee}$ for detectors CMD-3 and SND (VEPP-2000, 2 GeV), KLOE-2
(DA$\Phi$NE, 2.4 GeV), BES-III (BEPC\,II, 3.77 GeV), Belle (KEKB,
10.58 GeV), and BABAR (PEP-II, 10.58 GeV). They show that study of
scalar contributions in the $\gamma\gamma\to K^+K^-$ reaction near
the threshold can become quite wealthy at modern colliders
(currently, data on these contributions are absent).

\begin{table}\label{TAB1}
\caption{{\footnotesize Estimates of $N_{eeK^+K^-}$ for four values
of $E_{cm}$ and $L_{ee}$, and for two intervals of integration in
Eq. (\ref{NeeX}) $(\Delta\sqrt{s})_1$:
$\,2m_{K^+}<\sqrt{s}<1.05$\,GeV and $(\Delta\sqrt{s})_2$:
$\,2m_{K^+}<\sqrt{s}<1.1$\,GeV.}} \vspace{0.05cm}
\begin{center}
\begin{tabular}{|c|c|c|c|}
\hline $E_{cm}$  & $L_{ee}$       & $(\Delta\sqrt{s})_1$ & $(\Delta\sqrt{s})_2$  \\
\hline 2 GeV     & 1\,$fb^{-1}$   & $0.56\times10^3$ & $0.74\times10^3$ \\
\hline 2.4 GeV   & 5\,$fb^{-1}$   & $4.1\times10^3$  & $5.5\times10^3$  \\
\hline 3.77 GeV  & 5\,$fb^{-1}$   & $8.7\times10^3$  & $11.7\times10^3$ \\
\hline 10.58 GeV & 100\,$fb^{-1}$ & $5.1\times10^5$  & $6.9\times10^5$  \\
\hline\end{tabular}\end{center}\end{table}

The number of events in the $K^0_S\bar K^0_S$ channel can be
expected at least an order of magnitude smaller. But even
establishing a reliable upper limit on the $S$-wave $\gamma\gamma\to
K^0\bar K^0$ cross section near the threshold will be very important
for the selection of theoretical models that have simultaneously to
describe the reactions $\gamma\gamma$\,$\to$\,$\pi^+\pi^-$, $\gamma
\gamma$\,$\to$\,$ \pi^0\pi^0$, $\gamma\gamma$\,$\to$\,$\pi^0\eta$,
$\gamma\gamma\to K^+K^-$, and $\gamma\gamma\to K^0\bar K^0$ in the
$f_0(980)$ and $a_0(980)$ resonance region.

\vspace*{0.2cm}

This work was supported in part by RFBR, Grant No 10-02-00016, and
Interdisciplinary project No 102 of Siberian division of RAS.

\vspace*{0.4cm}


\end{document}